\documentclass[article]{aa}                   
\usepackage{graphicx}
\usepackage{natbib,txfonts}
\usepackage[colorlinks=true, citecolor=blue]{hyperref}
\usepackage{url}
\usepackage{soul}
\usepackage{changes}
\usepackage{xcolor}
\usepackage{orcidlink}

\def\kms{km\,s$^{-1}$}

\nolinenumbers

\begin{document}

\title{Transient narrow high-velocity absorptions in the stationary spectra of SS433}
\authorrunning{Dodin et al.}

   \author{A.V.~Dodin \orcidlink{0000-0002-6755-2120}\inst{1}\thanks{dodin\_nv@mail.ru}\,
   \and
   K.A.~Postnov \orcidlink{0000-0002-1705-617X}\inst{1,2}
   \and
   A.M.~Cherepashchuk \orcidlink{0000-0001-5595-2285}\inst{1}
   \and
   A.M.~Tatarnikov \orcidlink{0000-0002-4398-6258}\inst{1,3}
   }

   \institute{Sternberg Astronomical Institute, Lomonosov Moscow State University,
              13, Universitetskij prospekt, 119234 Moscow, Russia
              \and
              Kazan Federal University, 18 Kremlyovskaya st., 420008 Kazan, Russia
              \and
              Faculty of Physics, Lomonosov Moscow State University,  Leninskiye Gory 1-2, 119991 Moscow, Russia 
             }

   \date{Received ...; accepted ...}
    \abstract{
We report on the discovery of rare emergence (31 nights from 360 nights of observations) of narrow absorption features in hydrogen and helium lines in stationary SS433 spectra with velocities ranging from $-650$ to $-1900$ km~s$^{-1}$. The components arise independently of the appearance of P-Cygni line profiles which are frequently observed in the SS433 stationary spectra with terminal velocities ranging from $-200$ to $\sim -2500$~km s$^{-1}$. The characteristic rising time of the transient absorptions is about one day and the decay time is about two days. The phenomenology of the absorptions suggests their origin due to hydrodynamic instabilities of wind outflows from a supercritical accretion disk in SS433.}


   \maketitle
\section{Introduction}
\label{sec:intro}

SS433 is a Galactic microquasar -- a massive  eclipsing X-ray binary system at the advanced evolutionary stage with a precessing optically bright supercritical accretion disk and relativistic jets \citep{1979ApJ...230L..41M,1980ApJ...235L.131C,1981MNRAS.194..761C,2004ASPRv..12....1F,2020NewAR..8901542C}.

The system demonstrates three types of periodic variabilities: 
(1) precessional $P_{\rm prec}\approx 162.3^d$, (2) orbital $P_{\rm orb}\approx 13.1^d$ and (3) nutaional $P_{\rm nut}\approx 6.29^d$.
According to photometrical and spectral observations, these variabilities, as well as matter velocity in relativistic jets ($v_J\approx 80\,000$ km/s, are on average persistent over about 40 years \citep{2022ARep...66..451C}. The binary orbital period of SS433 shows evolutionary increase at a rate of $1.14\pm 0.25\times 10^{-7}$~s/s \citep{2021MNRAS.507L..19C} suggesting a high mass ratio of the binary components $q=M_x/M_v>0.8$ (here $M_x$ and $M_v$ are masses of the compact and optical component, respectively). Correspondingly, the distance between the components in SS433 secularly increases with time preventing the common envelope formation in the system \citep{2023NewA..10302060C} and agrees with evolutionary considerations for SS433 \citep{2017MNRAS.471.4256V}.

SS433 has been extensively observed across a wide wavelength range in radio, optical, X-ray and gamma-rays (see, e.g., \citealt{2020NewAR..8901542C}). Optical spectroscopic observations by different authors (see e.g. \citealt{2004ASPRv..12....1F}) reveled a lot of peculiarities in the spectrum of SS433. These peculiarities reflect the supercritical character of accretion on to the compact object in SS433 \citep{1973A&A....24..337S}. Observational evidences of the supercritical regime of accretion in SS433 include a huge optical luminosity of a geometrical thick accretion disk $\sim 10^{39}$~ erg/s \citep{1987SvA....31..295A}, as well as a powerful high-velocity disk wind outflow ($\dot M\sim 10^{-4} M_\odot$/yr) and the presence of relativistic jets.

A supersonic gas outflow from a supercritical accretion disk can be accompanied by various dynamical and thermal instabilities \citep{2024MNRAS.532.4826T}. These instabilities can manifest themselves both in variable integral radiation flux and spectral lines. It is well known (see, e.g., \citealt{2004ASPRv..12....1F}) that transient absorption P-Cyg components occur in blue wings of powerful stationary emission hydrogen and helium lines in SS433 suggesting the variable wind disk velocity from $\sim 500$ to $\sim 2000$ km/s. These changes correlate with the precessional period suggesting an anisotropic disk velocity outflow. 
It is interesting to search for additional appearances of the supercritical accretion regime in SS433 due to different accompanying  non-stationary phenomena.




During our spectral monitoring of SS433 with 2.5-meter telescope of the Caucasian Mountain Observatory (CMO) of SAI MSU \citep{2020gbar.conf..127S} we have discovered short-term emergence of isolated narrow absorption details in stationary hydrogen  lines with velocities typical for a  wind outflow from a supercritical accretion disk in SS433. Such absorptions can probe the innermost structure of the wind from the supercritical accretion disk and the outflow dynamics.  The fairly fable and transient nature of these features has rendered them unnoticed in previous studies of SS433 spectra. The goal of our letter is to draw attention to these features with non-typical for the outflow line profiles. 

\section{Observations}
 \label{sec:obs}
\begin{figure}
	\centering
	\includegraphics[width=0.85\linewidth]{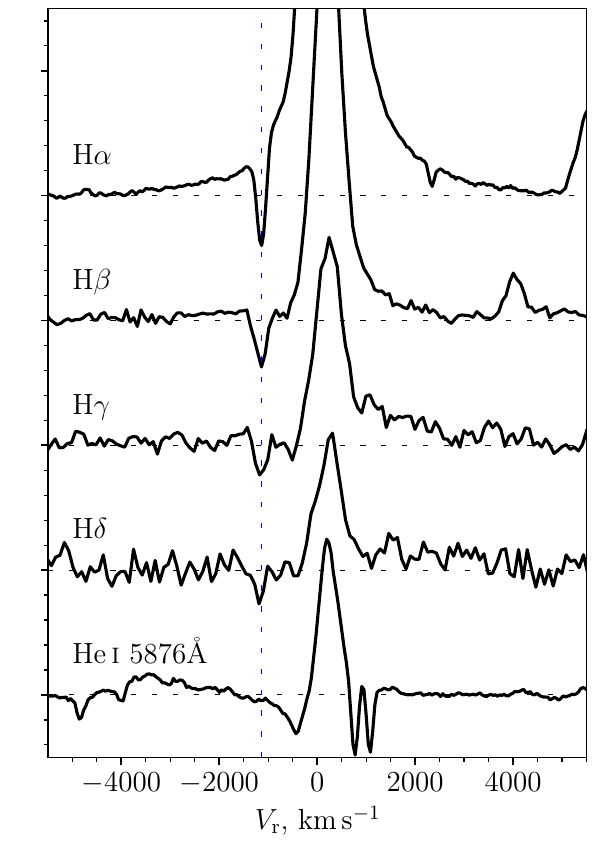}
	\caption{First discovery of narrow absorption component in the stationary TDS spectrum of SS433 on MJD 59741.87. Position of the absorption is marked with the dashed vertical line, the horizontal lines represent the continuum level. No absorption is visible in the He\,I\,5876{\AA} line. Interstellar lines are marked with red labels. The spectral resolution in H$\alpha$ and He\,I is about two times higher than in other lines.}
   \label{fig:discovery}
\end{figure}

\begin{figure*}
	\centering
	\includegraphics[width=0.75\linewidth]{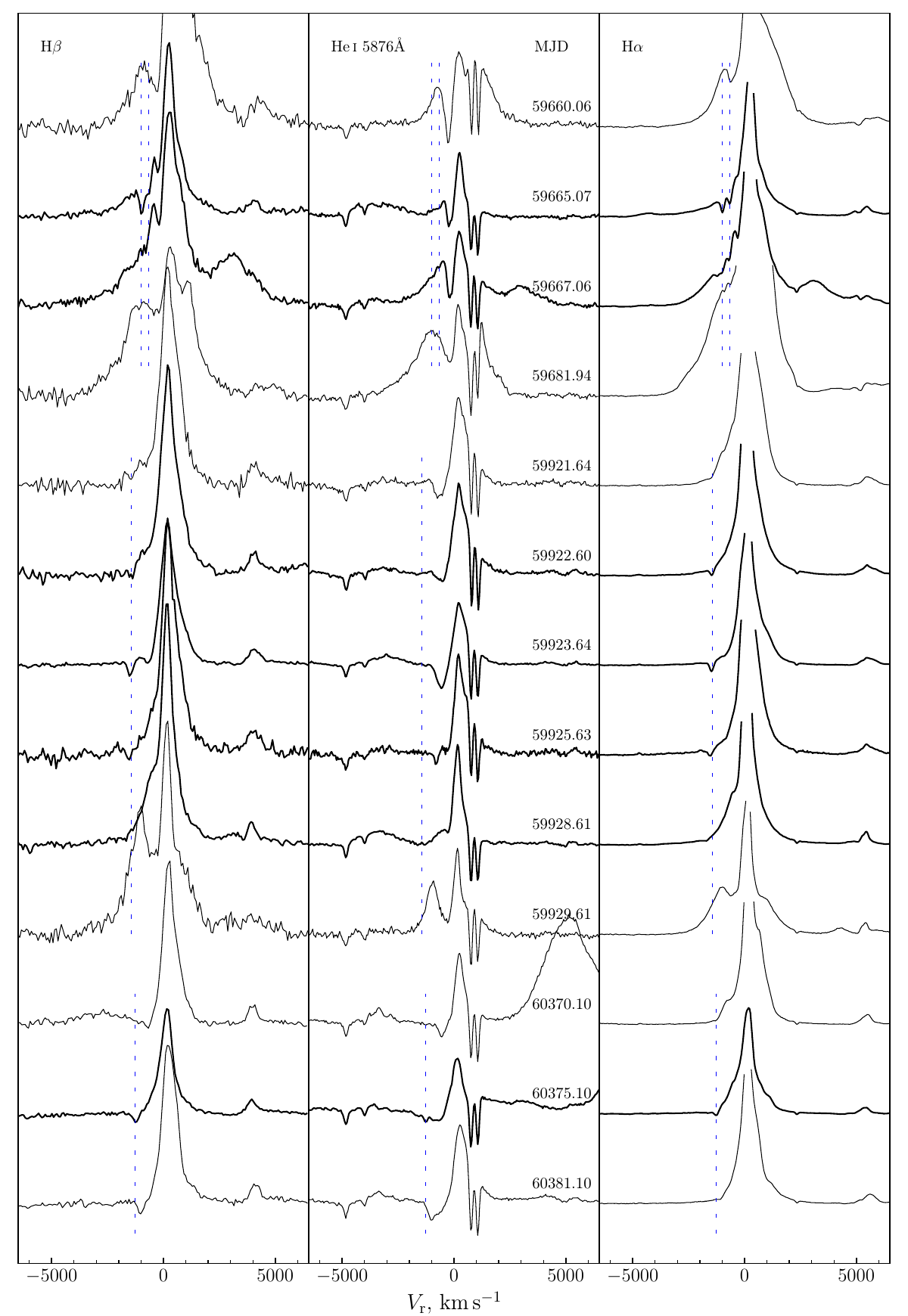}
	\caption{Three selected sequences of profiles of stationary H$\beta$, He\,I\,5876{\AA} and H$\alpha$ lines with transient absorptions: MJD 59660.06 -- 59681.94 with double narrow absorptions only in hydrogen lines; MJD 59921.64 -- 59929.61 with hydrogen absorptions including the rising and decay of the absorption; MJD 60370.10 -- 60381.10 with narrow absorption component in both hydrogen and helium lines. The vertical dashed lines mark the reference velocity of the absorption components. Red markers are for interstellar features.
 }
   \label{fig:anabs}
\end{figure*}

\begin{table*}
\caption{Parameters of the transient absorption. Different events are separated with horizontal lines.} 
\label{tab:anabs}
\tiny
\begin{tabular}{c | c | c | c | c | c | c | c | c | c}
\hline
MJD &$\varphi_{\rm Orb}$ & $\varphi_{\rm Prec}$& \multicolumn{5}{c|}{Equivalent width, \AA} & $V_{\rm r},$\,\kms& $\sigma,$\,\kms \\
\hline
         &  &  & H$\alpha$& H$\beta$& H$\gamma$& H$\delta$ & He\,I\,5876{\AA} &  &  \\
\hline
59130.81   & 0.176 & 0.249 & $0.97\pm0.05$ &  $1.70\pm0.05$  &  --  &  --      &    --       &   $ -1050\pm10$ & $71\pm4$ \\
59131.81   & 0.252 & 0.255 & $0.93\pm0.05$ &  $0.58\pm0.05$  &  --  &  --      &    --       &   $ -1024\pm8 $ & $90\pm6$ \\
59133.77   & 0.402 & 0.267 & $0.15\pm0.03$ &   --            &  --  &  --      &    --       &   $ -1115\pm8 $ & 0 \\
59134.78   & 0.479 & 0.273 & $0.48\pm0.03$ &   --            &  --  &  --      &    --       &   $ -1130\pm4 $ & $42\pm5$ \\
           &       &        & $0.25\pm0.02$ &   --           &  --  &  --      &    --       &   $  -922\pm4 $ & $<15$ \\
59135.73   & 0.552 & 0.279 & $0.52\pm0.04$ & $0.22\pm0.05$   &  --  &  --      &    --       &   $-1132\pm22$  & $54\pm7$ \\
           &       &       & $0.30\pm0.04$ &  --             &  --  &  --      &    --       &   $-958\pm12$   & $55\pm12$\\
\hline
59381.03   & 0.302 & 0.791 & $0.19\pm0.04$ &  --             &  --  &  --      &    --       &   $ -1516\pm8$  & $<38$ \\
59382.02   & 0.378 & 0.797 & $0.73\pm0.04$ &  --             &  --  &  --      &    --       &   $ -1510\pm4$  & $51\pm5$ \\
\hline
59559.62   & 0.954 & 0.891 & $1.21\pm0.05$ & $2.89\pm0.07$   & --   & --       &    --       &   $-1000\pm50$     &$89\pm3$ \\
\hline
59665.07   & 0.015 & 0.541 & $1.93\pm0.04$ & $1.28\pm0.14$   &  --  &  --      &    --       &   $-983\pm7$    & $62\pm2$ \\
           &       &       & $1.40\pm0.04$ & $0.59\pm0.13$   &  --  &  --      &    --       &   $-664\pm3$    & $60\pm3$ \\
59667.06   & 0.167 & 0.553  & $0.30\pm0.05$ &   --           &  --  &  --      &    --       &   $-1207\pm11$  & $52\pm17$ \\
           &       &        & $0.25\pm0.04$ &   --           &  --  &  --      &    --       &   $-1007\pm 9$  & <40 \\
           &       &        & $0.43\pm0.04$ &   --           &  --  &  --      &    --       &   $ -671\pm 5$  & <30 \\
           &       &        & $0.39\pm0.04$ &   --           &  --  &  --      &    --       &   $ -298\pm 5$  & <22 \\
\hline
59730.02   & 0.979 & 0.941 & $0.49\pm0.04$ & $0.28\pm0.04$   &  --  &  --      &    --       &   $ -946\pm 4 $  & $23\pm10$ \\
59730.97   & 0.052 & 0.947 & $0.70\pm0.04$ &   --            &  --  &  --      &    --       &   $ -1383\pm 4$  & $54\pm5$ \\
           &       &       & $0.82\pm0.04$ &   --            &  --  &  --      &    --       &   $  -966\pm 4$  & $45\pm4$ \\
59733.02   & 0.208 & 0.960 & $1.09\pm0.06$ & $1.03\pm0.07$   &  --  &  --      &    --       &   $ -999\pm20 $  & $99\pm5$ \\
\hline
59741.87   & 0.885 & 0.014 & $2.81\pm0.02$ &  $1.86\pm0.13$ & $1.03\pm0.23$  & $0.78\pm0.27$ & -- & $-1142\pm3$ & $73\pm1$ \\
\hline
59922.60   & 0.700 & 0.128 & $1.24\pm0.02$ &  $0.74\pm0.19$ &  $0.70\pm0.45$ &    --         & -- & $-1465\pm5$ & $57\pm2$ \\
59923.64   & 0.780 & 0.134 & $2.35\pm0.01$ &  $1.11\pm0.07$ &  $0.74\pm0.14$ & $0.18\pm0.18$ & -- & $-1478\pm3$ & $65\pm1$ \\
59925.63   & 0.931 & 0.147 & $2.02\pm0.05$ &  $0.56\pm0.24$ &      --        &     --        & -- & $-1535\pm4$ & $102\pm2$ \\
59928.61   & 0.159 & 0.165 & $0.29\pm0.02$ &       --       &      --        &     --        & -- & $-1703\pm5$ & $48\pm7$ \\
\hline
60137.97 & 0.163 & 0.455 & $0.32\pm0.02$  &   --          &    --          &     --          &   --           & $ -688\pm  4$ & $ 34\pm6  $ \\
         &       &       & $0.12\pm0.02$  &   --          &    --          &     --          &   --           & $ -946\pm  6$ & $  0      $ \\
60140.00 & 0.318 & 0.468 & $0.24\pm0.03$  &   --          &    --          &     --          &   --           & $ -451\pm  6$ & $  <26    $ \\
         &       &       & $0.57\pm0.04$  &   --          &    --          &     --          &   --           & $ -689\pm  5$ & $ 49\pm6  $ \\
60141.82 & 0.456 & 0.479 & $0.83\pm0.07$  & $1.05\pm0.32$ &    --          &     --          &   --           & $ -717\pm  5$ & $ 50\pm8  $ \\
60147.95 & 0.926 & 0.517 & $0.27\pm0.03$  &   --          &    --          &     --          &   --           & $ -833\pm  6$ & $ 53\pm10 $ \\
60151.00 & 0.159 & 0.535 & $0.36\pm0.05$  &   --          &    --          &     --          &   --           & $ -867\pm  8$ & $ 47\pm12 $ \\
         &       &       & $0.15\pm0.06$  &   --          &    --          &     --          &   --           & $-1243\pm 23$ & $ 60\pm30 $ \\
60151.87 & 0.225 & 0.541 & $0.33\pm0.05$  & $0.48\pm0.09$ &    --          &     --          &   --           & $ -804\pm  6$ & $ 34\pm10 $ \\
\hline
60375.10 & 0.289 & 0.916 & $1.48\pm0.08$  & $0.88\pm0.11$ & $0.54\pm0.12$  &   $0.43\pm0.13$ &  $0.31\pm0.03$ & $-1252\pm  4$ & $ 78\pm4  $ \\
\hline
60537.83 & 0.728 & 0.919 & $0.31\pm0.03$  &   --          &   --           &     --          &  $0.29\pm0.03$ & $ -980\pm  5$ & $  <15    $ \\
60543.76 & 0.181 & 0.955 & $0.28\pm0.03$  &   --          &    --          &     --          &   --           & $ -885\pm  6$ & $ 50\pm10 $ \\
         &       &       & $0.60\pm0.05$  &   --          &    --          &     --          &   --           & $-1281\pm  7$ & $120\pm10 $ \\
60544.83$^a$ & 0.263 & 0.962 & $0.78\pm0.09$  & $0.33\pm0.04$ &  $0.38\pm0.10$ &     --          &  $0.50\pm0.01$ & $ -880\pm  6$ & $ 46\pm10 $ \\
\hline
\multicolumn{10}{l}{$^a$The transient absorption is also visible in the He\,I 7065{\AA} line with the equivalent width of $0.19\pm0.04${\AA}.}
\end{tabular}
\end{table*}

Our observations are based on spectral monitoring of SS433 which has been carried out on the 2.5-m telescope of CMO SAI MSU since the end of 2019 using the Transient Double-beam Spectrograph (TDS). The spectral resolution of the TDS is $R=\lambda/\Delta \lambda \approx 2400$ in the red channel $(0.56 - 0.74$~$\mu$m) and $\approx 1300$ in the blue $(0.36 - 0.56$~$\mu$m) (see \citealt{Potanin20} for a detailed description of TDS and data reduction). In addition to the standard processing procedure, a telluric correction was performed using A0V stars with known synthetic spectrum. In total, we have obtained around 340 individual spectra (one spectrum per night). In addition to our spectral monitoring, we used the archive data obtained with X-shooter spectrograph \footnote{\url{https://archive.eso.org/scienceportal}} \citep{2011A&A...536A.105V} during another 23 nights. Telluric lines in the X-shooter spectra have not been removed but are marked where necessary.

\section{Transient narrow absorptions in stationary spectra of SS433}

During our TDS observations of SS433 in 11.06.22 (MJD 59741.87) we  came across an unusual spectrum with pronounced narrow absorptions in the stationary hydrogen emission lines with a velocity of about -1150 {\kms} that falls below the continuum level. The absorptions were observed in all stationary hydrogen lines  H$\alpha$ -- H$\delta$ with sufficient signal-to-noise ratio (see Fig. \ref{fig:discovery}).
Unfortunately, the sky conditions made it impossible to determine the moment of neither emergence nor disappearance of this absorption. However, further observations and the scrutinizing of all available TDS spectra of SS433 enabled us to find other nine similar events. In Fig.\,\ref{fig:anabs} we show examples of three episodes of the appearance of the absorption. Parameters of the lines for all detected events are listed in Table\,\ref{tab:anabs}. The transient absorptions appear in hydrogen lines, but sometimes weak absorption emerge in He\,I lines 5876{\AA}, 7065{\AA}
(in Table\,\ref{tab:anabs} we show only results for the He\,I\,5876{\AA} line).

\begin{figure}
	\centering
	\includegraphics[width=\linewidth]{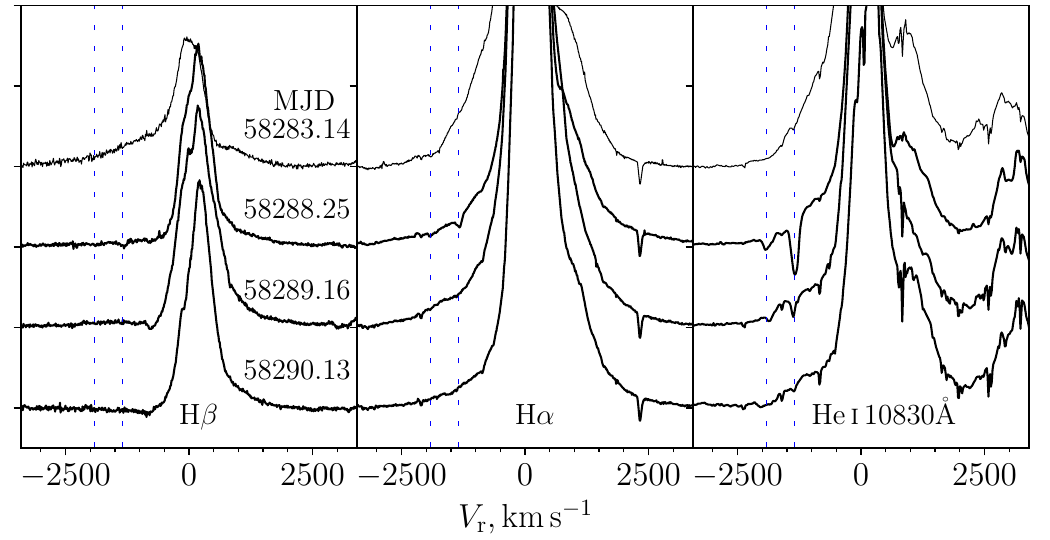} \label{tab:xshoo}
	\caption{The sequence of H$\beta$, H$\alpha,$ He\,I\,10830{\AA} profiles in the X-shooter spectra with emerging transient absorptions. The vertical dashed lines as in Fig.\,\ref{fig:anabs}. The narrowest spectral features near He\,I 10830{\AA} are telluric lines, examples are marked with $\oplus.$}
   \label{fig:xshoo}
\end{figure}

A similar absorption was found to appear in the archive X-shooter spectra (MJD 58288.25 -- 58289.16, see Table\,\ref{tab:Xshooter}). In the X-shooter spectra the absorption is present in hydrogen lines but absent in helium lines in the visual band. The absorption is especially strong in the He\,I\,10830{\AA} line (see Fig. \ref{fig:xshoo}), which, unlike visual He lines, is formed from a metastable level. 

The properties of the transient absorptions are summarized as follows.
\begin{enumerate}
\item  The absorption has a narrow width slightly exceeding the instrumental profile. The width of a Gaussian instrumental profile is  $\sigma_{\rm TDS}=52$ {\kms} in the H$\alpha$ \citep{2023AstBu..78..283B} for a full slit filling by stellar light. To evaluate the absorption width we have quadratically subtracted the instrumental width $\sigma_{\rm TDS}$ from the observed one  $\sigma_{\rm Obs}$: $\sigma^2=\sigma^2_{\rm Obs} - \sigma^2_{\rm TDS}$ (see Table\,\ref{tab:anabs}). Table\,\ref{tab:anabs} suggests that the typical line width is about  $\sigma=50$\,\kms. If the line width $\rm FWHM\sim100$ {\kms} is due to angular spread of a shell radially moving with a constant velocity of 1000 km\,s$^{-1},$ the subtended semi-angle of the absorbing region should be around $25$ degrees.

\item The high velocity of the absorption $\sim1000$ {\kms} with the FWHM $\sim100$ {\kms} is distinct from usual P-Cygni absorptions where the line width is comparable to the terminal velocity of the outflow. The transient absorption looks like an isolated spectral line in the far blue wing of the H$\alpha$ emission or on top the continuum for other Balmer hydrogen lines
(see Figs.\,\ref{fig:discovery}, \ref{fig:anabs}). 

\item The profile does not display a separate emission component that could be related to the same gas as that producing the absorption. The point is that in a P-Cygni profile, absorption is produced only by the gas that is projected on the background source while the rest of the gas forms emission. The emission strength depends on the source function in the gas, its area and optical depth. 

\item The absorption in our TDS spectra is observed predominantly in hydrogen lines. The X-shooter spectra also shown the absorption in hydrogen lines, but the strongest absorption is found in the quasi-resonant He\,I\,10830{\AA}. Only in two out of 11 episodes the transient absorption was observed in the neutral helium lines 
He\,I\,5876{\AA} and He\,I\,7065{\AA}. The usual P-Cygni-like absorption in SS433 is observed both in hydrogen and helium, with helium absorption being more pronounced due to smaller contribution of the central wide emission. 

\item The characteristic rising time of the absorption is about one day, the fading time is about two days (see the episode  MJD = 59922.60 -- 59928.61 in Table \ref{tab:anabs}). The absorbing gas moving with a velocity of $\sim1000$ {\kms} passes a distance of about one a.u. over this time interval. Assuming that the line width is due to velocity gradients in the shell, the gas would expand by $\sim2.35\sigma \Delta t\sim 0.1$ a.u. per day. 

\item Equivalent widths of the transient H$\beta$ and H$\alpha$ narrow absorption lines are comparable, whereas in the optical thin case the  H$\beta$ width should have been 10 times as small as the H$\alpha$ width. This means that the  H$\alpha$ line has a larger optical depth. 

\item There can be several such absorption details with different velocities in one observation (see Fig.\,\ref{fig:anabs} MJD=59665, 59667).

\item In December 2022 (MJD 59922.6-59928.6), the absorption velocity monotonically (parabolically) increased with time after the moment of appearance in the spectrum from 1465 {\kms} to 1700 \kms.

\item The absorption can emerge simultanelously with the usual P Cygni absorption (see MJD=59923.64 and other profiles in Fig.\,\ref{fig:anabs}).
\end{enumerate}

\begin{table}
\caption{Parameters of the transient absorption in the X-shooter spectra}
\label{tab:Xshooter}
\begin{tabular}{ c | c | c | c }
\hline
Line& EW, \AA & $V_{\rm r},$ \kms & $\sigma,$ \kms \\
\hline
\multicolumn{4}{c}{MJD = 58288.25, $\varphi_{\rm Orb}=0.771,$  $\varphi_{\rm Prec}=0.057$}  \\
    H$\beta$     & $0.12\pm0.02$ & $-1329\pm  5$  & $  55 \pm  6$   \\
    H$\beta$     & $0.08\pm0.02$ & $-1882\pm 22$  & $ 108 \pm 26$   \\
    H$\beta$     & $0.06\pm0.02$ & $ -774\pm  9$  & $  50 \pm 10$   \\
    H$\alpha$    & $0.24\pm0.02$ & $-1333\pm  2$  & $  48 \pm  3$   \\
    H$\alpha$    & $0.17\pm0.02$ & $-1883\pm  8$  & $  80 \pm  9$   \\
    H$\alpha$    & $0.04\pm0.01$ & $ -883\pm  9$  & $  33 \pm 11$   \\
    He\,I\,10830{\AA}   & $3.13\pm0.03$ & $-1335\pm  1$  & $  66 \pm  1$   \\
    He\,I\,10830{\AA}   & $0.85\pm0.05$ & $-1904\pm  3$  & $  73 \pm  3$   \\
    He\,I\,10830{\AA}   & $0.27\pm0.05$ & $ -769\pm 11$  & $  76 \pm 11$   \\
\multicolumn{4}{c}{MJD = 58289.16, $\varphi_{\rm Orb}=0.840,$  $\varphi_{\rm Prec}=0.063$}  \\    
  He\,I\,10830{\AA}   & $0.69\pm0.03$ & $-1367\pm  2$  & $  46 \pm  2$   \\
  He\,I\,10830{\AA}   & $0.41\pm0.04$ & $-1841\pm  4$  & $  57 \pm  5$   \\
\multicolumn{4}{c}{MJD = 58290.13, $\varphi_{\rm Orb}=0.914,$  $\varphi_{\rm Prec}= 0.069$}  \\      
 He\,I\,10830{\AA}   & $0.23\pm0.03$ & $-1353\pm  4$  & $  40 \pm  5$   \\
 He\,I\,10830{\AA}   & $0.72\pm0.07$ & $-1924\pm  8$  & $ 134 \pm 10$   \\
\hline
\end{tabular}
\end{table}

\section{Discussion}
 \label{sec:results}

The observed absorption line width FWHM $\sim100$\,{\kms} 
cannot be explained by the stream line divergence of a spherically expanding shell in the SS433 wind because at distances above one a.u., a cloud covering the continuum source $\sim 0.3$ a.u. in size would expand in a conical region with an opening semi-angle of 10 degrees, instead of the required 25 degrees. Thus, the line stream divergence in a radial wind could explain only a half of the observed line width. 

Internal velocity gradients in the line formation region seem to be more plausible mechanism of the line broadening. The lack of helium lines but with rare exceptions (see Table \ref{tab:anabs}) evidence for a relatively cold gas producing the absorption. We have carried out a series of \textsc{Cloudy}  simulations \citep{2023RMxAA..59..327C} suggesting that absorptions with the required properties could be formed in a wide range of the gas parameters with a temperature of about $10\,000$ K and a small fraction of the ionizing radiation. Higher temperatures would give rise to the appearance of helium absorption lines. 

\begin{figure}
	\centering
	\includegraphics[width=1\linewidth]{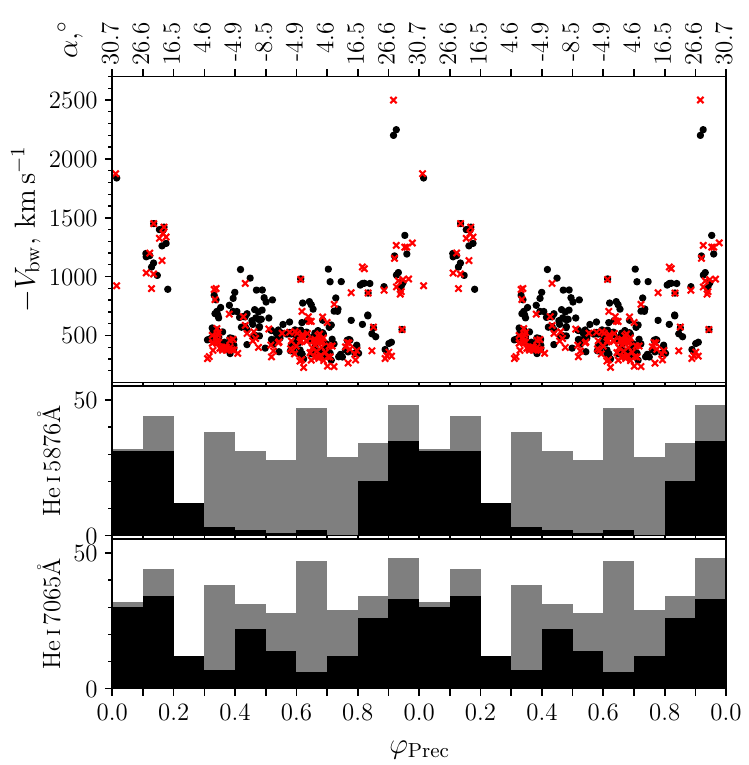}
	\caption{The upper panel: terminal velocities of the blue wing of  P-Cygni profiles as a function of the precessional phase. On the top, the viewing angle $\alpha$ of the disk plane at the corresponding precessional phases is shown. 
 Points and crosses mark the He\,I\,5876{\AA} and   He\,I\,7065{\AA} lines, respectively. The bottom panels show histograms of distributions of $\varphi_{\rm Prec}$ of all observed spectra (in gray) and those without P-Cygni profiles of stationary emissions in SS433 spectra (in black).} 
   \label{fig:pcyg}
\end{figure}

The observed velocities of the transient absorption fall within the range of the disk outflow velocities derived from P-Cygni line profiles in SS433. \citet{Fabrika1997} found that disk outflow velocity increases with increasing angle above the accretion disk. We have checked this statement using our data and confirmed that the terminal  velocity of the blue wing of P-Cygni absorptions can attain $\sim2500$ {\kms} during the maximum opening disk angle in SS433 (see Fig.\,\ref{fig:pcyg}, upper panel). It is seen also in Fig. \ref{fig:pcyg} (middle and bottom panels) that blue P-Cygni absorptions are most frequently observed with velocities from 500 to 1000 {\kms} at phases far from the maximum disk opening. On the other hand, in the maximum disk opening phases the P-Cygni absorptions are rarely observed, and if appear, they demonstrate high velocities up to $\sim 2500$ \kms. 

We believe that the appearance of 
cold gas condensations on the line of sight can be due to the Kelvin-Helmholtz instability in sheared wind outflow leading to internal shocks, compression and rapid gas cooling. Numerical simulations of supercritical disk accretion  \citep{2024MNRAS.532.4826T} also predicts large-scale turbulent motions in the outflow above the disk plane. The cold gas condensation leaving the line of sight and their fragmentation can be the reason for rapid disappearance of the transient absorptions. The observed episode of transient absorption's radial velocity increase could be related to a shock wave propagation along the gas with decreasing density because the velocity of the contact discontinuity (the region where the cooling gas is accumulated behind the shock front) is determined by the balance between the ram and gas pressure in the contact discontinuity frame. 

\begin{figure}
	\centering
	\includegraphics[width=1\linewidth]{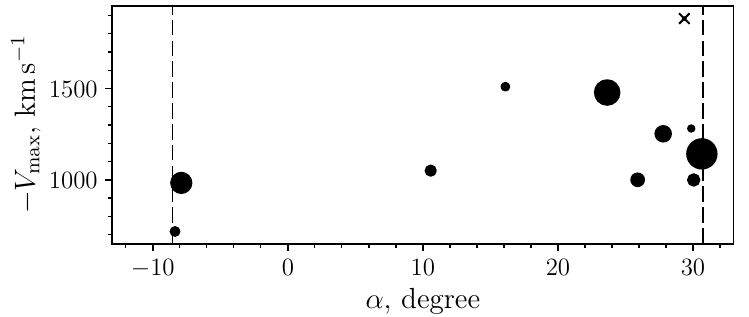}
	\caption{The maximum velocity in the episode with a transient absorption emerging as a function of the disk inclination angle to the line of sight (zero corresponds to the edge-on disk view). The size of the filled circles is proportional to the absorption equivalent width in H$\alpha.$ The cross marks the X-shooter observation. The dashed lines correspond to the limiting angles of the disk inclination angle according to the kinematic model of jets in SS433 \citep{1981ApJ...246L.141H,2018ARep...62..747C}.}
   \label{fig:absapp}
\end{figure}

If our hypothesis is true and the transient absorptions arise due to internal disk wind instabilities then, as in the case of usual P Cygni absorptions, we should expect the relation with the precessional period of the disk. More precisely, such absorptions should be found more frequently and with higher velocities during the precessional phases with the maximal disk opening to the observer.  In Fig. \ref{fig:absapp} we show the dependence of the velocity of the absorptions and their equivalent width on the angle between the line of sight and the disk plane. It is clear that indeed the absorptions are never observed when the disk is seen edge-on and tend to concentrate to the maximum disk opening to the observer. 


\section{Conclusion}
 \label{sec:conclusion}

Our long-term spectroscopic monitoring of SS433 has revealed  comparatively rare episodes during which  narrow absorption components in the blue wing of the stationary emission H$\alpha$ line emerge and disappear on the characteristic time of several days .
The appearance of such narrow transient absorptions is most likely related to the supercritical accretion regime and evidences for specific non-stationary phenomena in the supercritical accretion disk outflow. The detected transient spectral features are relatively fable and rare and could have been missed or considered as artifacts in the stationary spectra of SS433. 

The increase in the statistics of the emerging of such absorptions in stationary spectra of SS433 and study of their properties will help to investigate non-stationary physical processes in the supercritical accretion disk wind. Therefore, further careful spectral observations of SS433 seems to be very promising. 


\section*{Acknowledgements}

AVD and AMCh are supported by RSF grant 23-12-00092 (observations and data processing, participation in the interpretation and discussion of the results). Observations with telescopes of Caucasian Mountain Observatory of SAI MSU are supported by the Program of development of M.V. Lomonosov Moscow State University.


\bibliographystyle{aa}  
\bibliography{ss433}    

\end{document}